\documentclass[11pt,twoside]{article}
\usepackage{newpasp}
\usepackage{epsf}

\newcommand{\hgpcC}{{h^{-3}\rm\,Gpc^3}}

\newcommand{\hmpc}{{h^{-1}\rm\,Mpc}}

\newcommand{\ihmpc}{{h\rm\,Mpc^{-1}}}

\begin{document}
\title{Large-Scale Structure \& Future Surveys}
\author{Daniel J.\ Eisenstein}
\affil{Steward Observatory, 933 N. Cherry Ave., Tucson AZ, 85721}

\begin{abstract}
As the 2dF Galaxy Redshift Survey and Sloan Digital Sky Survey move
toward completion, it is time to ask what the next generation of
survey of large-scale structure should be.  I discuss some of the 
cosmological justifications for such surveys and conclude that 
surveys at $z=3$ offer a critical advantage in their ability to
access linear-regime clustering at scales smaller than any current
survey and even the CMB.  I discuss a possible implementation of
such a survey and highlight some of the potential science return.
\end{abstract}

\section{Introduction}

The study of perturbations in the universe plays
a central role in modern cosmology.  Not only do these
perturbations create the opportunity for objects like galaxies
and clusters to form, but they also record a history of
the early phases of the universe.  By studying the detailed
statistics of the perturbations imprinted on the 
cosmic microwave background (CMB) or on the galaxy distribution,
we can learn about the time scales and matter content of the
universe and perhaps even about very early physics, such as
the nature of inflation.  

Over thirty years of galaxy surveys have produced an ever-improving
map of the density structure of the universe.  With the 
Sloan Digital Sky Survey (SDSS; York et al.\ 2000) and 2-degree Field
Galaxy Redshift Survey (2dFGRS; Colless et al.\ 2001), we will have over
a million galaxy redshifts tracing the density in the local universe.
These surveys will measure the 
clustering of galaxies on huge scales and will allow detailed
comparison of the clustering properties of different subclasses
of galaxies on intermediate and small scales.
Current surveys at higher redshift, for example, the CNOC II
(Yee et al.\ 2000), DEEP (http://deep.ucolick.org/),
and VIRMOS (http://www.astrsp-mrs.fr/virmos/vvds.htm)
surveys, plan to acquire around 100,000 redshifts
of galaxies from much earlier epochs.  This will allow 
us to measure the evolution of galaxy properties, including
their clustering, over cosmic time.

Large redshift surveys have been among the heaviest users of our 
wide-field multi-object spectroscopic capabilities.  As the current
generation of surveys moves towards completion, it is very timely
(indeed, past due) to begin planning for the next.  There are several
different options, not all of which involve large amounts of spectroscopy.
In my contribution, I discuss some of the options and offer a sketch of 
one possibility, a large redshift survey at $z\approx 3$.  

\section{What can we learn from large-scale structure?}

In planning for the future, we must begin with what we hope to learn
from surveys of large-scale structure.  
This discussion breaks into two sets of topics.
First, large-scale structure depends upon, and therefore can 
inform, fundamental cosmology.  
The current state of perturbations in the universe is a product of
both the initial seeding of the fluctuations and their evolution
throughout cosmic history.  These two ingredients are usually
separable, especially when combined with CMB anisotropy measurements.
The spatial dependences of clustering statistics can
measure the abundances and properties of cosmic matter, e.g.,
the baryon fraction, neutrino mass, and the coldness of the dark matter,
as well as the spectrum of the initial fluctuations, e.g.,
the spectral tilt or perhaps any intrinsic scales.
The time evolution of clustering statistics further
depends on cosmological parameters, e.g. the matter density of the universe
or the properties of dark energy.  Large-scale structure can also
be sensitive to non-Gaussianity in the initial seeding of perturbations.
Finally, measurements of peculiar velocities and redshift distortions
test the role of gravity in driving the evolution of the cosmic perturbations.

Second, with regard to galaxy surveys, one is probing the properties
of galaxy clustering bias.  While this is often seen as a nuisance
in the context of large-scale structure, measurements of clustering bias
are an opportunity to place constraints on 
the theory of galaxy formation.  With large surveys, we are 
moving beyond the idea of bias as a single number to investigate
the dependences of bias on spatial scale, time, and 
intrinsic properties of the galaxies.  These dependences should
encode the relation of galaxies both to mass and to other galaxies.
Clustering bias should be a precision test of galaxy formation theory.
However, it is not clear that current theory can utilize the
precision of even the current generations of surveys, and so 
I will focus my discussion strictly on cosmological inferences
from large-scale structure.

It is important to recognize the value of the ``linear regime'', 
i.e.~the smallness of fluctuations on large scales.  
In the theory of gravitational instability, small fluctuations
evolve in simple ways, namely that the evolution of each Fourier mode 
depends on its wavelength and on the homogeneous properties of the 
universe but remains independent from all other perturbative modes.  
This means that linear-regime perturbations retain full
memory of their history.  Perturbations on smaller scales, in
which the rms fractional density fluctuation is order unity or larger,
grow in a non-linear fashion in which all modes become coupled to
one another.  This erases many imprints of the early universe 
(e.g., Meiksin et al.~1999).
Hence, searches for preferred scales or subtle non-Gaussian signatures 
in the early universe are generally only possible on large, linear scales.

Some of the cosmological signatures available in linear-regime 
clustering include: 
\begin{itemize}
\item Features in the initial power spectrum.  The initial spectrum
is usually assumed to be a power-law in wavenumber, as is common
in inflationary models.  However, there are models that predict
more complicated spectra, and there has been persistent albeit
controversial observational evidence for a peak in the power spectrum
on scales of $120\hmpc$ (e.g., Broadhurst et al. 1990).
\item Baryon acoustic oscillations.  Prior to $z\approx1000$,
the coupling of the photons and electrons in the ionized universe
causes perturbations to act as sound waves (Peebles \& Yu 1970).  
The rapid loss of 
pressure support at recombination captures a snapshot of these
oscillations.  The result is harmonic signatures in the CMB
and matter power spectra.  The effect is smaller in the matter
spectrum because the baryons are subdominant to the cold dark matter,
which does not participate in the acoustic motion.
\item Neutrino masses.  Massive neutrinos suppress clustering on small
scales relative to large scales because they move too quickly to
be trapped in small fluctuations (Bond \& Szalay 1983).
\item Initial non-Gaussianity.  Most inflation models produce
Gaussian initial conditions for the density perturbations, but
there are exceptions.  If primordial non-Gaussianity exists, then
the linear regime carries even more information and
would be an important new window onto the process that initiates
structure formation.
\end{itemize}

Fortunately, clustering bias may also be simple in the linear regime.  
Models in which galaxy formation is a local (i.e., sub-Mpc) process
predict a scale-independent bias on large scales (Coles 1993).  
This is especially
secure when seeking preferred scales, e.g. the baryon acoustic oscillations,
as it is implausible that galaxy formation would involve action on 
such enormous and specific scales.

In the present-day universe, the linear regime conservatively would
be defined as wavelengths above $60\hmpc$ or wavenumbers $k<k_{NL}=0.1\ihmpc$.
Sharp spatial features, such as the baryon acoustic oscillations,
are erased on smaller scales (Meiksin et al. 1999).  
At higher redshift, clustering is
less advanced and so the linear regime extends to smaller comoving
scales.  However, because $\Omega_m\sim0.3$, the non-linear scale recedes
only slowly at low redshifts.  At $z\approx1$, the non-linear scale
is at best a factor of two smaller ($k_{NL}=0.18\ihmpc$).
At redshifts above 1, the non-linear scale begins to recede quickly,
both because of the faster evolution and because of the shape
of the CDM power spectrum on small scales.  At $z\approx3$, the
non-linear scale is roughly 5 times smaller than present ($k_{NL}=0.5\ihmpc$).

\section{Surveys in Cosmological Context}

Measuring clustering in the linear regime requires enormous survey
volumes with moderate, but not superb, sampling densities.  The 2dFGRS
survey will provide about a survey volume of about $0.1\hgpcC$.
The SDSS main galaxy survey is currently of similar size and
will eventually triple this volume.  The SDSS luminous, red
galaxy (LRG; Eisenstein et al. 2001) 
sample will probe a yet larger volume, about $1\hgpcC$.

Current surveys at $z\approx1$ probe considerably less volume
than this, which means that they do not compete well in measuring
linear regime clustering.  These surveys are of course
wonderful for studying galaxy evolution, including the change
in clustering on intermediate scales.
Surveys of quasars have too small a number density
to sample the density field properly.  

It is worth noting that increasing the volume of a ``local'' sample 
by a factor of 10 over SDSS LRG would require a $\pi$ steradian survey 
to $z=1$ with at least a million galaxies.  Volume alone drives
us to high redshift!

Weak lensing surveys are quickly maturing (see Wittman 2002 for a review).  
This method carries great 
promise because it avoids the
problems of galaxy bias.  The ability to measure mass fluctuations
directly will be a great advance on both linear and non-linear
scales.  In particular, weak lensing studies will measure the 
amplitude, tilt, and smoothed shape of the matter power spectrum
to excellent precision (e.g., Hu \& Tegmark 1999).  
Note that this applies to both $z<1$
clustering, using galaxies as the sources, and $z\approx 3$ clustering,
using the CMB as the source, although the latter is limited to $k<0.2\ihmpc$
even for optimistic future experiments (Hu 2001, Hu \& Okamoto 2002).
It will be difficult for galaxy
clustering surveys to compete on these topics, because the amplitudes
and broad tilts are exactly the items with which clustering bias 
interferes.

Nevertheless, weak lensing in cosmological contexts does have 
some weaknesses.  In particular, because sources are lensed by all
mass along the line of sight, weak lensing suffers heavily from projection 
effects.  This is crippling for the study of sharp features in the
power spectrum and initial non-Gaussianity.  Separating the lensing
signal by source redshift provides a small amount of resolution along
the line of sight but not enough for these topics (Hu 1999).

The primary anisotropies of the CMB (those produced at $z\approx1000$) 
are also the product of the linear regime.  
The Planck mission (http://astro.estec.esa.nl/SA-general/Projects/Planck/)
should be able to measure the primary
signal to $\ell\approx2500$, which is $k\approx 0.25\ihmpc$.
Converting the available precision to the
requirements on a galaxy redshift survey 
produces an equivalent survey volume
of $20\hgpcC$!  This is of course superb.
However, on smaller scales, the primary anisotropies of the 
CMB will be very difficult to measure.  

\section{Surveys at $z=3$}

I have argued that linear-regime clustering is a path to
fundamental cosmology.  How do we improve the coverage
described in the last section?  The scale of survey required
to improve the current situation at $k<0.1\ihmpc$, much less
to compete with the CMB, suggests that we consider how to
access the linear regime at $k>0.2\ihmpc$.
For this, we must work at higher redshift, and the next convenient
window for optical spectroscopy is $z\approx3$ using the Lyman break
selection technique (Steidel et al. 1996).  
Let us therefore consider a large survey at $z\approx 3$.

As a primary purpose of such a survey is to search for preferred
scales, i.e., sharp features in the power spectrum, it would
be best to design a survey that can certainly detect the one
such feature that we strongly expect to exist, namely the baryon
acoustic oscillations.  
The first oscillation appears at $k<0.1\ihmpc$
and should be accessible in current low-redshift surveys (Percival et al. 2001 
claims a 2-$\sigma$ hint of the oscillations).  The second oscillation
is still available at $z<1$, but the current mid-redshift surveys 
are probably too small for the expected amplitude.  A survey at
$z=3$ could recover the entire sequence, limited only by the sharply
declining amplitude of the higher harmonics.

Detecting the third and fourth acoustic peaks\footnote{As a matter of
nomenclature, one should note that the peaks in the matter 
power spectrum appear half as often as the peaks in the CMB power spectrum.  
Hence, the fourth peak in the matter power spectrum is (roughly) the 
seventh peak in the CMB power spectrum.} 
would require a survey at $z\approx 3$ of roughly $0.5\hgpcC$ at
the optimal number density of tracers.  The latter is set by the
inverse of the power spectrum on the desired scale (Kaiser 1986).  For Lyman
break galaxies, which are known to have $\sigma_8\approx 1$ (Steidel et al. 1998), the
desired sampling is roughly $0.001h^3{\rm\;Mpc^{-3}}$, which is
(coincidentally and fortunately) roughly the available number of 
targets accessible to few-hour exposure times on 8-meter class telescopes.
The total number of galaxies in the survey would be half a million,
and they would be distributed over about 150 square degrees with a 
sampling a 1 galaxy per square arcminute.

\begin{figure}[p]
\plotone{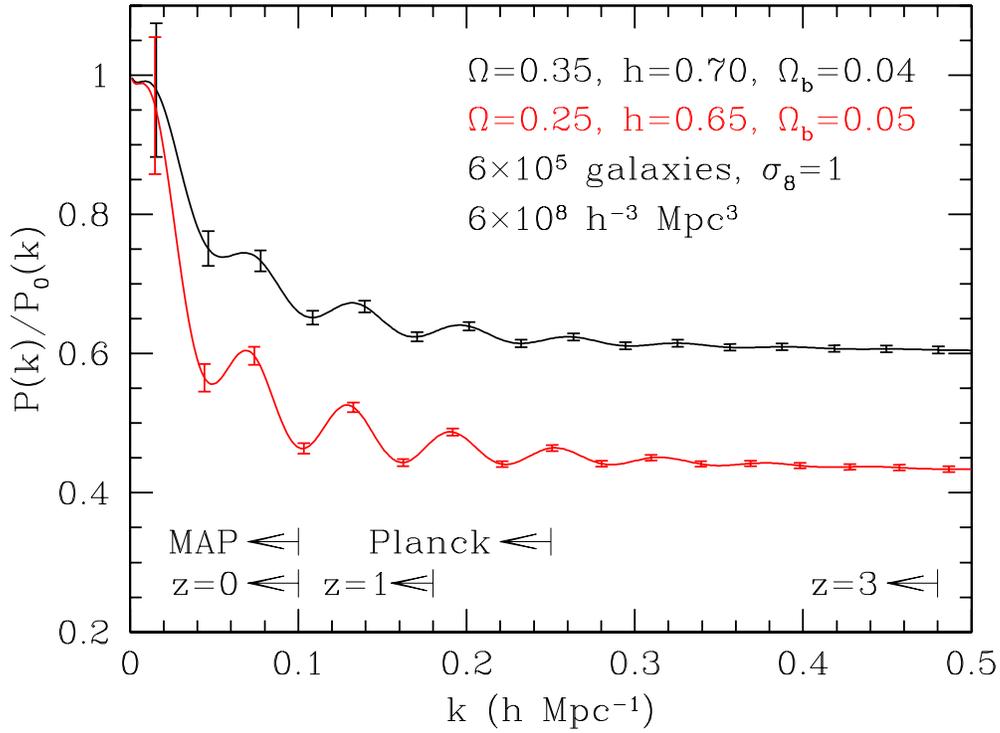}
\caption{\label{bumps}%
The statistical error bars on a large redshift survey at $z=3$
superposed on the power spectra of two different cosmological models.
The power spectra have been divided by the zero-baryon power spectra
of their $\Omega_m$ and $H_0$.  Both cosmological models have the
usual baryon density of Big Bang nucleosynthesis.  The top curve has
$\Omega_m=0.35$ and $H_0=70$ km/s/Mpc, while the bottom curve has
$\Omega_m=0.25$ and $H_0=65$ km/s/Mpc.
The baryon acoustic oscillations are the wiggles in both power spectra.
We have assumed 600,000 galaxies spread over $0.6\hgpcC$.  The galaxies
are assumed to have a clustering amplitude of $\sigma_8=1$.  The survey
would detect 3 or 4 of the acoustic oscillations, depending on cosmology,
with a marginal detection of one additional peak.  The arrows show the
range of the linear regime at different redshifts and for two CMB
experiments, MAP and Planck.}
\end{figure}

The statistical precision on the power spectrum available to such a 
survey is shown in Figure \ref{bumps}, following the approximations of
Tegmark (1997).  The survey would yield 1\%
measurements of bandpowers narrow enough to sample the acoustic
peaks.  It would detect 3 or 4 of the peaks, depending on cosmology,
with a marginal detection of one additional peak in each case.

What value does the detection of acoustic peaks have, given that
we (likely) have detected them in the CMB power spectrum?
The key idea is that the scale of these oscillations defines 
a standard ruler.
At low redshift, this allows us to split the
angular diameter distance degeneracy of the CMB, thereby measuring the
Hubble constant (Eisenstein et al. 1999).  
At higher redshift, we actually get to measure 
this distance twice, once along the line of sight and once transverse
to the line of sight.  This cosmological distortion is familiar
to those using the Alcock-Paczynski (1979) test, but here we need not rely
on the ratio of the distances, as the distance itself is known from
the shape of the acoustic peaks in the CMB.  Hence, one is able to 
constrain both the angular diameter distance to $z=3$ and the Hubble
constant at $z=3$.  Clearly, this will put leverage on dark energy.
In the limit that the dark energy is a cosmological constant, then
the Hubble constant at $z=3$ is very nearly $\sqrt{\Omega_m H_0^2}(1+z)^3$.
This measurement of $\Omega H_0^2$ would allow a consistency check
with the CMB, the failure of which could be resolved by permitting
extra relativistic species at $z\approx 10^4$.

The exact value of such a measurement in the context of joint parameter
estimation with CMB, supernovae distance-redshift relations, weak lensing
experiments, and other probes of classical cosmology remains to be
calculated.

Lest one become too focused on baryon oscillations, 
it should be stressed that the fine structure of the linear regime 
on the scales $0.25\ihmpc<k<0.5\ihmpc$ is terra incognita to any current or
planned survey, yet this is exactly the regime that controls the
formation of large galaxies.  
The level of precision indicated in Figure \ref{bumps}
would open a new discovery space for effects in the initial
power spectrum and its subsequent evolution.

\section{Why spectroscopy?}

An alternative to a large redshift survey at $z=3$ would be an
even larger imaging survey using photometric redshifts 
to create the angular correlations of Lyman-break 
galaxies\footnote{To be clear, the photometric redshifts are not
of sufficient precision to create an effectively 3-dimensional survey; rather
one is simply creating a thinner slab of space within which to do
2-dimensional analyses.}.  
A proposal similar to this, but at lower redshift, 
has been made by Cooray et al. 2001.
Such a survey would access the linear regime in a similar fashion to the
discussion in the previous section.  

It is worth noting some of the disadvantages of this approach.  
First, the imaging survey would have to be considerably larger to match
the precision, several thousand square degrees instead of 150.  
Second, the projection would significantly reduce leverage on 
non-Gaussianity and slightly reduce sensitivity to narrow features,
such as the higher acoustic oscillations.  Third, one loses the 
ability to measure the line-of-sight distance of the acoustic peak
and other Alcock-Paczynski effects.  
Finally, such a survey offers less control over systematic errors.  
All the modes in a single-band imaging survey would be strictly angular, 
meaning that they are degenerate with angular errors in the map 
(e.g., photometry offsets).  
A spectroscopic survey can remove such errors by using modes 
that are mixed between radial and tangential directions (Tegmark et al. 1998).  
With photometric redshifts, an imaging survey does have some redundancy,
namely that different slabs should be uncorrelated, but this is partially
compromised by the scatter in photometric redshifts and by the need
to control systematic errors in multiple bands of imaging.
Given that we seek precision measurements, the redundancy
of the spectroscopic survey is a notable advantage.

\section{Conclusions}

I have described some of the landscape in the design of the next
generation of large-scale structure surveys.  The linear regime 
remains a very powerful window into cosmic history.  Recognizing
this value, the most persuasive extension beyond current surveys 
is to seek smaller scales rather than additional precision on large scales.
This can be done only by observing the universe at higher redshift.
$z\approx3$ is the next plausible window.  A large survey of around half
a million $z\approx3$ galaxies over 150 square degrees would probe
the linear-regime power spectrum to scales twice as small as the CMB
primary anisotropies with percent level precision.
This is sufficient to detect four acoustic oscillations, which in
turn can be used to measure the Hubble constant and angular diameter
distance to $z=3$.  We must quantify in detail how such a survey
would compare to and complement other possible cosmological measurements.

A survey of this size and scope is challenging but not impossible.
It is probably too large for current wide-field spectrographs, but
one can certainly imagine next-generation instruments that could 
perform the observations in reasonable time.  If the science case can
be shown to be quantitatively compelling, then such a survey would
be a feasible next step in cosmology.

\end{document}